\documentclass[12pt]{article}
\let\useblackboard=\iftrue
\useblackboard
\font\blackboard=msbm10 scaled \magstep1
\font\blackboards=msbm7
\font\blackboardss=msbm5
\newfam\black
\textfont\black=\blackboard
\scriptfont\black=\blackboards
\scriptscriptfont\black=\blackboardss
\def\Bbb#1{{\fam\black\relax#1}}
\else
\def\Bbb{\bf}
\fi
\newcommand{\divergence}{{\log L\over 2\pi}}
\newcommand{\Bars}{{\rm (Bars)}}

\newcommand{\ka}{\kappa}

\newcommand{\BR}{\Bbb{R}}

\newcommand{\half}{{1\over 2}}
\newcommand{\ha}{{1\over 2}}

\newcommand{\RR}{{\cal R}}

\newcommand{\be}{\begin{equation}}
\newcommand{\ee}{\end{equation}}
\newcommand{\ben}{\begin{eqnarray}\displaystyle}
\newcommand{\een}{\end{eqnarray}}
\newcommand{\refb}[1]{(\ref{#1})}
\newcommand{\p}{\partial}
\newcommand{\ov}{\over}
\newcommand{\CB} {{\cal B}}
\newcommand{\CT} {{\cal T}}
\newcommand{\Tr}{{\rm Tr}}
\newcommand{\sectiono}[1]{\section{#1}\setcounter{equation}{0}}

\begin{document}
{}~ \hfill\vbox{\hbox{hep-th/0202087}\hbox{CTP-MIT-3240}
\hbox{RUNHETC-2002-03} }\break

\vskip 1.3cm

\centerline{\large \bf
Open String Star as a Continuous Moyal Product
}

\vspace*{8.0ex}

\centerline{\large  Michael R.~Douglas$^{a,b,c}$, Hong Liu$^a$, Gregory Moore$^a$
 and Barton Zwiebach$^d$}

\vspace*{6.0ex}

\centerline{\large \it ~$^a$Department of Physics }

\centerline{\large \it Rutgers University, Piscataway, NJ 08544,
USA}

\centerline{E-mail: \{mrd,gmoore,liu\}@physics.rutgers.edu}

\vspace*{2.8ex}

\centerline{\large \it ~$^b$I.H.E.S., Bures-sur-Yvette 91440 France}

\vspace*{2.8ex}

\centerline{\large \it ~$^c$ Isaac Newton Institute for Mathematical
Sciences,}
\centerline{\large \it Cambridge, CB3 0EH, U.K.}

\vspace*{2.0ex}

\centerline{\large \it $^d$Center for Theoretical Physics}

\centerline{\large \it
Massachussetts Institute of Technology,}

\centerline{\large \it Cambridge, MA 02139, USA}

\vspace*{1.2ex}

\centerline{E-mail: zwiebach@mitlns.mit.edu}

\vspace*{5.0ex}

\centerline{\bf Abstract}
\bigskip
We establish that the open string star product
in the zero momentum sector
can be described as a continuous tensor product of
mutually commuting two dimensional Moyal star products.
Let the continuous variable $\kappa \in [~0,\infty)$
parametrize the eigenvalues of the Neumann
matrices; then the noncommutativity parameter is given
by $\theta(\kappa) =2 \tanh ({\pi\kappa\over 4})$. For each
$\ka$, the Moyal coordinates are a linear combination
of even position modes, and the Fourier transform of a
linear combination of odd position modes. The commuting
coordinate at $\ka=0$ is identified as the momentum
carried by half the string.  We discuss
the relation to Bars' work, and attempt to write the string
field action as a noncommutative field theory.

\vfill \eject

\baselineskip=16pt

\tableofcontents

\sectiono{Introduction and summary}\label{s0}

The star product of open string  functionals is
a noncommutative associative product based on
a simple prescription for the gluing of the underlying open
strings~\cite{OSFT}. In this prescription, the right-half of the
first string must coincide with the left-half of the second string,
and the resulting string is composed by the left-half of the first string
together with the right-half of the second string. As with
pointwise multiplication of functions, where position space delta
functions define the kernel imposing the coincidence conditions, the open
string star product uses delta functionals of half-strings to impose the
requisite coincidence conditions. The language of delta functionals,
however, is difficult to use in explicit computations and the methods
of conformal field theory were used to develop Fock space
descriptions of the star product as a three string
vertex~\cite{gross-jevicki,cremmer}.
While these concrete constructions have allowed much recent progress
in open string field theory, at a
foundational level important questions remain. The conditions that
specify the space of open string fields for which the axioms of the algebra
hold are not yet known. Moreover, the star algebra has not been
given a structural definition in terms of well-understood algebras.

\medskip
In a stimulating paper Bars \cite{0106157} set out to describe
the open string star product in terms of the Moyal product.
This product is the unique associative deformation (up to isomorphism) of the
pointwise multiplication of functions on $\BR^{2n}$,
and as such has played a central role in recent studies and
constructions of noncommutative
field theories (see, for example \cite{0012145,0102076,0106048}).
Bars proposed that each even position mode (except for the zero mode)
together with a specific linear combination of the Fourier transforms of
the odd position modes forms a Moyal pair. The various Moyal pairs
are mutually commuting and they all have the same noncommutativity
parameter $\theta\not=0$.
The precise treatment of the
center-of-mass or string midpoint was left for future work.

In this paper, we investigate the description of the open string
star product in terms of Moyal products using a different starting
point. Recently, the Neumann matrices defining the oscillator
form of the three-open-string vertex have been diagonalized and
the spectrum and eigenvectors have been constructed explicitly
\cite{0111281}. In addition, orthonormality and completeness of
the eigenvectors was proven in \cite{0201015}.  These results
imply that there is a basis of oscillators where the exponential
in the vertex takes
diagonal form, and thus furnishes a representation of the product
in terms of mutually commuting algebras.

We find that each commuting factor in this product is a Moyal product,
and thus each  algebra is a (one dimensional) Heisenberg algebra. Since
the spectrum of the Neumann matrices is continuous, we get a continuous
tensor product of Heisenberg algebras with a
smoothly varying noncommutativity
parameter. It should be emphasized that we work on the subspace
of zero momentum functionals, that is, functionals that are
independent of the center of mass coordinate of the string.

\medskip
Our Moyal coordinates are constructed as follows.
For  each $\kappa \in
(-\infty,\infty)$  the Neumann
matrices $M, M^{12},$ and  $M^{21}$ have a common eigenvector $v(\kappa)$ with
eigenvalues $\mu(\kappa), \mu^{12}(\kappa)$ and $\mu^{21}(\kappa)$
respectively. The signal of noncommutativity is the fact that
$\mu^{12}(\kappa)
\not= \mu^{21}(\kappa)$. Under the action of open string twist, the
eigenvectors $v(\ka)$ and $v(-\ka)$  mix, and one can conveniently define
twist odd and twist even eigenvectors for $\ka\in [0,\infty)$. The
unpaired eigenvector at $\ka=0$ is actually twist odd, and has featured
in several studies of vacuum string field theory and
sliver states~\cite{0111034,0111069,0201136,0111129}. It will have
an important role here as well. The definite twist combinations are
degenerate eigenvectors of $M$, and while they not eigenvectors of $M^{12}$
and
$M^{21}$, these matrices have simple action on them.
We find that for each $\ka >0$, the twist even and the twist odd
eigenvectors respectively define the Moyal coordinates $x(\ka)$ and
$y(\ka)$, with  noncommutativity parameter given 
as\footnote{With units 
inserted back there is an $\alpha'$ factor in the right hand side of \refb{thetres}.
So, as usual, $\theta$  has dimensions of (length$)^2$.}
\be
\label{thetres}
\theta(\ka) = 2 \,\tanh \Bigl( {\pi\ka\over 4} \Bigr) \,,\quad \ka\geq 0\,.
\ee
Note that for $\kappa=0$, where there is only one eigenvector, and
thus only one coordinate, there is no scope for a Moyal product, and
consistently the
noncommutativity vanishes. The various coordinates for different values of
$\ka$ commute, so that we have
\be \label{defthk}
[  x(\ka) , y(\ka') \, ]_* = i\, \theta(\ka) \, \delta(\ka - \ka')\,.
\ee

\bigskip
The description of the star product in terms of the above
Moyal products is as follows.
Given a string field $\Psi (X(\sigma))$, we view  it
as a function $\Psi (\{ x_{2n}\} , \{
x_{2n-1}\})$ of the even and odd modes in the expansion of $X(\sigma)$.
Then we  Fourier transform $\Psi$ on the odd modes
$x_{2n-1}$ into variables $p_{2n-1}$ (that correspond to the eigenvalues
of $\hat p_{2n-1}$) obtaining a functional
$\widetilde \Psi (\{ x_{2n}\} , \{ p_{2n-1}\})$. This
functional is now writen in
terms of coordinates $(x(\ka) , y(\ka))$  using the invertible relations
\ben
\label{thenewvar0}
x (\ka) =\,\,\sqrt{2} \,\,\sum_{n=1}^\infty  v_{2n}(\ka)
\sqrt{2n}\, \,
 x_{2n}\,,\qquad
 y(\ka) = -\sqrt{2} \sum_{n=1}^\infty  {v_{2n-1}(\ka)
\over\sqrt{2n-1}}  \,\, p_{2n-1}\,,
\een
that, as mentioned before, use the twist even and twist odd eigenvectors.
The resulting field
$\Psi^M ( x(\ka) , y(\ka))$ is star multiplied just using the
Moyal product on the underlying coordinates. \footnote{ 
Strictly speaking, the two products are related by multiplication
by an overall (infinite) constant factor; see sections 4 and 5.}
As we will see, this continuous Moyal product can be written
as a {\it functional} Moyal product, using the language of
path integrals over the coordinates $(x(\ka), y(\ka))$.
For $\ka=0$ the surviving coordinate $y(\ka=0)$ is predicted
to be commutative. As can be seen in \refb{thenewvar0} it corresponds
to a specific linear combination of momentum modes. This combination
precisely selects the momentum carried by half of the string. As we
explain in the text, for zero momentum functionals, the open string
vertex treats the momentum carried by half of the string as a
commuting coordinate.

Alternatively, we can invert~\refb{thenewvar0} and use~\refb{defthk}
to interpret the Witten star product as the Moyal product in the noncommutative
space $(x_{2n}, p_{2m-1})$ with  a nondiagonal form of noncommutative
parameter
\be \label{xpcom0}
[x_{2n},p_{2m-1}]_{\ast} = i  \Theta^{2n,2m-1} \qquad n,m\geq 1 \,.
\ee
where  the
matrix $\Theta$ above coincides with one of the matrices relevant
in the half-string formalism.
In this way Witten's prescription for identifying half strings may be
interpreted as identifying the end points of dipoles in the noncommutative
space~\refb{xpcom0}. It is interesting that $p_{2m-1}$ have the interpretation
of coordinates and thus $x_{2m-1}$ should be interpreted as momenta.

We will compare  in detail our results with those in the earlier work of Bars
\cite{0106157} finding  broad agreement, but some subtle differences. His
result,  given in terms of Moyal pairs involving the original modes, is
indeed compatible with~\refb{xpcom0}.
The subtleties concerning  the
invertibility of this matrix are presumably responsible for the
absence of the above mentioned commuting coordinate in the framework of
\cite{0106157}.

We have also explored the possibility of rewriting open string field
theory in the language of NC field theory. For this we simplify the
string field theory action by simply ignoring the ghost sector, by
working at zero momentum and 
in a fixed gauge. The
kinetic operator is then $L_0$ and a kernel representing its action on
the continuous basis of oscillators is required and seems to exist,
albeit in a regulated form. It is then possible to rewrite the $L_0$
action as star commutation. Our final results are suggestive, but are
clearly of very preliminary nature.

This paper is organized as follows.  In section \ref{s1} we
explain how the basic two-dimensional Moyal product can be
described as a three-vertex in a Fock state formalism. In section
\ref{s2}, we introduce the continuous basis of oscillators needed
to diagonalize the Neumann matrices and rewrite the open string
vertex in terms of them. The comparison between sections \ref{s1}
and \ref{s2} is done in section \ref{s3} where we identify the
Moyal structures in the star algebra, and give the physical
interpretation for the commuting coordinate. In section~\ref{s4}
we give  a functional description of the continuous tensor product
of the Moyal algebra. In section \ref{s5} we compare our results
with those of Bars, and discuss some issues related to the half
string formalism. In section \ref{s6} we attempt to rewrite open
string field theory in the language of noncommutative field 
theory.  We offer some concluding remarks in section \ref{s7}.

A remark on notation.  Although our results can of course be interpreted
as defining the Witten string field theory product as the multiplication
law of a noncommutative algebra, for clarity we will not take this step,
instead writing the noncommutative product explicitly as ``$f * g$.''

\sectiono{Oscillator vertex for the  Moyal product} \label{s1}

The three string vertex is ordinarily expressed  as a quadratic form
in oscillators acting on the vacuum. More precisely, we have oscillators
from the three state spaces, those of the two input string fields and
that of the output product, acting on the tensor product of vacua.
In order to understand how to interpret such a vertex as a Moyal
product we will calculate here an oscillator-form three-vertex for
the Moyal multiplication in two dimensional space.

It is useful first to recall how ordinary commutative pointwise
multiplication of functions can be encoded  in a three-vertex written
in oscillator form. This result was given in \cite{0111069} (eqn. (3.15)):
\ben
\label{compr}
|V_3\rangle = \Bigl({2\over 3\sqrt{\pi}}\Bigr)^{1/2} \exp \Bigl[\,\,
{1\over 6}
(a_1^\dagger a_1^\dagger  +a_2^\dagger a_2^\dagger+a_3^\dagger a_3^\dagger
 )
 -{2\over 3} \,\,
(a_1^\dagger a_2^\dagger +a_2^\dagger a_3^\dagger + a_3^\dagger a_1^\dagger
 )\Bigr] |0\rangle\,,
\een
where the subscripts one and two on the oscillators refer to
the first function and second function to be multiplied respectively,
while the subscript three is associated to the product. This
result goes along with the definitions
\be
\label{zmodepos}
\hat x = {i\over \sqrt{2}} (a- a^\dagger)\,, \qquad  \hat p =  {1\over
\sqrt{2}} (a+ a^\dagger)\,,
\ee
and the construction of the position states
\ben
\label{xef}
\langle x| = {1\over \pi^{1/4}}\langle 0| \exp \Bigl( - {1\over
2}
x^2 + \sqrt{2}\,\, i  a \,  x + {1\over 2}
a a \Bigr)\,.
\een
Given two functions $f(x) = \langle x | f\rangle$ and
$g(x) = \langle x | g\rangle$, then $ (f\cdot g)(x) =
\langle x | \langle f| \langle g| |V_3\rangle$.  We are after
the generalization of \refb{compr} for the case of Moyal multiplication.
It should be noted from the outset that for a Moyal product we need
two coordinates -- so the vertex will be a modification of the
above involving two sets of oscillators.  When the noncommutativity
is set to zero, we must recover two copies of the above vertex.

\bigskip
\noindent
We begin with the standard definition of the Moyal product in momentum
space
for $\BR^{2d}$:
\be
\label{moyaldef}
(f * g) (k) =
\int {dk_1\over (2\pi)^{2d}} {dk_2\over (2\pi)^{2d}}\,\,
(2\pi)^{2d} \delta(k_1+ k_2 - k)
\, e^{-{i\over 2} k_{1\mu} \Theta^{\mu\nu} \, k_{2\nu} } \,
f(k_1) \,g(k_2) \,.
\ee
with Fourier transformation definitions
\be
h(k) = \int dx  \, e^{ik x} h(x)\,, \quad h(x) = \int {dk\over
(2\pi)^{2d}}
 \, e^{-ik x} h(k)\,.
\ee We pass to coordinate space to find a kernel $K(x_1,x_2,x_3)$
defined from \be (f * g) (x_3) \equiv \int dx_1 dx_2 \,
K(x_1,x_2,x_3) \, f(x_1) g(x_2)\,. \ee Using the formula for
momentum space we have: \be K(x_1,x_2,x_3) = \int {dk_1\over
(2\pi)^{2d}}{dk_2\over (2\pi)^{2d}} \, \exp\Bigl( - {1\over 2}\,
k_{1}\cdot  i\Theta \cdot\, k_{2}  + k_1\cdot  i(x_1-x_3) + k_2
\,\cdot i(x_2-x_3) \Bigr)\,.
 \ee
Here $\Theta$ is a $2d\times 2d$
matrix and we use $\cdot$ to denote scalar product, or sum over
$2d$ component indices. The integral can be done by completing
squares and one finds the standard result \be \label{defkf}
K(x_1,x_2,x_3) = {1\over \pi^{2d} \det \Theta}  \exp \Bigl ( -2i
\,(x_1-x_3)
 \Theta^{-1}\, (x_2- x_3 ) \Bigr)\,,
\ee
or, in manifestly cyclic form
\be
\label{ee1}
K(x_1,x_2,x_3) =  {1\over \pi^{2d} \det \Theta}  \exp \Bigl ( -2i\, [
x_1
 \Theta^{-1}\, x_2 + x_2 \,\Theta^{-1} \, x_3 + x_3 \,
\Theta^{-1} \, x_1 ]\, \Bigr)\,.
\ee

When $\Theta$ is skew diagonal, it is possible
write the star product in a mixed momentum and coordinate basis and
the kernel takes form of a product
of $\delta$-functions~\cite{bigatti,0106157}. More explicitly
write
\be
\vec x = (x^M, y^M)   \,,
\ee
where $x^M, y^M \in \BR^{d}, \, M=1, \cdots d$ and
\be \label{theblo}
[x^M, y^N] = i \theta^{MN}\,.
\ee
Now consider a mixed basis $(x^M, p_N^{(y)})$ and define
\be \label{lrco}
x_L = x^M + \ha \theta^{MN} p_N^{(y)}, \qquad
x_R = x^M - \ha \theta^{MN} p_N^{(y)}\,.
\ee
Then the Moyal product can be written as
\be \label{moymix}
(f \ast g) (x_L, x_R) =
{1 \ov (4 \pi)^d \sqrt{\det \Theta}} \int d^d z \, f (x_L, z) g (z, x_R) \,.
\ee
The above representation has a nice physical interpretation~\cite{bigatti}
in terms of dipoles interacting by joining their ends together.
The center of the dipole
is specified by the center of mass coordinates $\vec x$.  The momenta
$\vec p = (p_M^{(x)}, p_N^{(y)}) $
defines the  extent of the dipole as  $\vec \Delta = \Theta \cdot \vec p$.
Then $x_L, x_R$ are coordinates for the left and right end of the
dipole.\footnote{Since the dipole endpoints
are noncommutative, we cannot simultaneously
identify the other coordinates $(y_L, y_R)$ when
joining the dipoles in the star product.}

\medskip
\noindent
Now restrict attention to  two dimensional
noncommutative space, i.e., set $d=1$.
Write a general cyclic ansatz for a three-vertex, by considering
oscillators $(a,b)$ associated to the coordinates $x^\mu$, with $\mu=
1,2$.
\be
(x_1, y_1\, ;  x_2, y_2\, ;  x_3, y_3\,) \,
\leftrightarrow \,(  a_1^\dagger, b_1^\dagger\,;\,  a_2^\dagger,
b_2^\dagger\,;\,
 a_3^\dagger, b_3^\dagger)\,.
\ee
The position eigenstates can be defined using \refb{zmodepos} together with
the analogous relations
\be
\label{zmodeposmore}
 \hat y =
{i\over
\sqrt{2}} (b-  b^\dagger)\,; \qquad   \hat q =  {1\over \sqrt{2}} (b+
b^\dagger)\,.
\ee
Using 2-component notation
$
\vec x = (x,y),  \,\, \vec a = (a, b)$,
we have
\ben
\label{xrefpref}
\langle \vec x| &= {1\over \sqrt{\pi}}\langle 0| \exp \Bigl( - {1\over
2}
\vec x \cdot
\vec x + \sqrt{2}\,\, i \vec a \cdot \vec x + {1\over 2}
\vec a\cdot \vec a \Bigr)\,,
\nonumber\\
\langle \vec p|&= {1\over \sqrt{\pi}}\langle 0| \exp \Bigl( - {1\over 2}
\vec p \cdot
\vec p + \sqrt{2}\,\,  \vec a \cdot \vec p - {1\over 2} \vec a\cdot \vec
a
\Bigr)\,.
\een
The three-vertex will be
written as
\be
|V_3\rangle  =
N\exp \Bigl( -{1\over 2} \vec A^\dagger \, V \, \vec A^\dagger \Bigr)\,
|0\rangle\,,
\ee
where $V$ is a $6\times 6$ matrix, and $\vec A$ a six component vector
\be
\label{vmat}
V = \pmatrix{ u&v&v^T \cr v^T &u&v \cr v&v^T&u } \,,\quad
\vec A^\dagger = (\vec a^\dagger_1,\vec a^\dagger_2,
\vec a^\dagger_3 )\,.
\ee
with $u,v$ two by two matrices.   We now require that
\be
\label{meq}
K(\vec x_1,\vec x_2,\vec x_3) = \Bigl( \langle \vec x_1|\otimes
\langle \vec x_2|\otimes\langle \vec x_3|\Bigr) \, |V_3\rangle
= {1\over \pi^{2} \det \Theta}  \exp \Bigl (-{1\over 2}
\vec X\, K \, \vec X \Bigr)\,,
\ee
where use was made of \refb{ee1}, and we have defined
\be
 K = (2i) \pmatrix{ 0&\Theta^{-1}&-\Theta^{-1} \cr -\Theta^{-1}
 &0&\Theta^{-1} \cr \Theta^{-1} &-\Theta^{-1} &0 }\,, \quad
\vec X = (\vec x_1,\vec x_2,
\vec x_3 ) \,.
\ee
In this notation,  \refb{xrefpref} allows one to write
\be
 \langle X | = \langle \vec x_1|\otimes
\langle \vec x_2|\otimes\langle \vec x_3| =
 {1\over \pi\sqrt{\pi}}\langle 0| \exp \Bigl( - {1\over 2} \vec X \cdot
\vec X + \sqrt{2}\,\, i \vec A \cdot \vec X + {1\over 2} \vec A\cdot
\vec A \Bigr)\,.
\ee
Substituting back into  the main condition \refb{meq} and doing the
contraction we find:
\be  {1\over \pi^{2} \det \Theta}  \exp \Bigl (-{1\over 2}
\vec X\, K \, \vec X \Bigr) ={N\over \pi\sqrt{\pi}}
{1\over \sqrt{\det (1+ V)}} \exp\Bigl( -{1\over 2} \vec X\,
{1-V\over 1+V}\, \vec X\Bigr)\,,
\ee
and conclude that
\be
 V= {1-K\over 1+K} \,, \qquad
N = {\sqrt{\det (1+ V)}\over  \sqrt{\pi} \det \Theta }\,.
\ee
We now take
$\Theta = \pmatrix{0 &\theta \cr
-\theta &0}$, and
find that
for $\theta \not=0$ the matrix $(1+K)$ appearing in $V$ is invertible.
The resulting expression for $V$ in terms of the $u,v$ matrices in
\refb{vmat} is
\be
u = {-4 + \theta^2\over 12 + \theta^2}\pmatrix{1&0\cr 0 &1}\,,\quad
v = {4 \over 12 + \theta^2} \pmatrix{2&i\theta \cr -i\theta &2}\,.
\ee
Moreover
\be
N  = {8\over \sqrt{\pi} (12+ \theta^2) } = {2\over 3\sqrt{\pi}}\,\,
{1\over \Bigl(1  + {\theta^2\over 12}\Bigr)}\,.
\ee
All in all, the form of the vertex is
\ben
\label{canmoy}
|V_3(\theta)\rangle &=&{2\over 3\sqrt{\pi}}\,\, {1\over 1
+ {\theta^2\over 12}}\, \exp \Bigl[ - {1\over 2}
 \Bigl( {-4 + \theta^2\over 12 + \theta^2}\Bigr)
(a_1^\dagger a_1^\dagger
+ b_1^\dagger b_1^\dagger + \hbox{cyclic} ) \nonumber \\
&&\qquad\qquad\qquad\quad -\Bigl( {8\over 12+ \theta^2} \Bigr) \,\,
(a_1^\dagger a_2^\dagger + b_1^\dagger b_2^\dagger + \hbox{cyclic} )\\
&&\qquad\qquad\qquad\quad -\Bigl( {4i\theta\over 12+ \theta^2} \Bigr)
\,\,
(a_1^\dagger b_2^\dagger -  b_1^\dagger a_2^\dagger  
+ \hbox{cyclic} )\Bigr] |0\rangle \,.\nonumber
\een
This is the desired form of the vertex for canonical Moyal. The limit
$\theta \to 0$ is smooth and gives:
\ben
|V_3(\theta=0)\rangle = {2\over 3\sqrt{\pi}}\, \exp \Bigl[\,\,  {1\over
6}
(a_1^\dagger a_1^\dagger + b_1^\dagger b_1^\dagger +
\hbox{cyclic} )
 -{2\over 3} \,\,
(a_1^\dagger a_2^\dagger +
b_1^\dagger b_2^\dagger + \hbox{cyclic} )\Bigr] |0\rangle\,.
\een
Happily, this agrees
with the commutative
result reviewed in \refb{compr}.

\sectiono{Open string star in the continuous oscillator basis}
\label{s2}
\medskip

In this section we will examine the three string vertex and use
the recent diagonalization of the Neumann matrices to rewrite
the vertex in a basis of oscillators with a continuous mode label.
We will then perform a redefinition of the oscillators in such
a way to allow a comparison with the Moyal form of the vertex
determined in the previous section.

We begin by recalling the main results of \cite{0111281} that we will
need.
We have the
set of eigenvectors of
$v_m (\ka)$ of matrices
$M^{rs}_{mn}$  which are labeled by a continuous parameter $-\infty<
\kappa< \infty$, i.e.
\be
\label{eveqn}
\sum_{n=1}^{\infty} M_{mn}^{rs} v_n (\ka) = \mu^{rs} (\ka) v_m (\ka),
\ee
where the eigenvalues are given by
\ben
\label{evexp}
\mu (\ka) &=&\mu^{11} (\ka) = - {1 \over 1 + 2 \cosh {\pi \ka \over 2}},
\nonumber \\
\mu^{12} (\ka) &=&{1+ \cosh {\pi \ka \over 2} +
\sinh {\pi \ka \over 2}  \over 1 + 2 \cosh {\pi \ka \over 2}}, \\
\mu^{21} (\ka) &=&{1+ \cosh {\pi \ka \over 2} - \sinh {\pi \ka
\over 2}  \over 1 + 2 \cosh {\pi \ka \over 2}}\,,\nonumber
\een
and
they satify the relations \be \mu + \mu^{12} + \mu^{21}=1, \qquad
\mu^{12}\mu^{21}= \mu(\mu-1) \ . \ee The eigenvector $v_n (\ka)$
is  given by the generating function:\footnote{
The  $v_n$'s  here differ from those in~\cite{0111281}
by the inclusion of the normalization factor
$N(\ka)^{-\ha}$.}
 \be
\label{genaf} f_{\ka} (z) = \sum_{n=1}^{\infty} {v_n (\ka) \ov
\sqrt{n}} z^n = {1 \ov N (\ka)^{\ha}} {1 \ov \ka} (1 - e^{- \ka
\tan^{-1} z})\,,
 \ee 
where $N (\ka)$ is given by~\cite{0201136}
 \be \label{defnorm}
 N (\ka) = {2 \ov \ka} \sinh {\pi \ka \ov 2} \,.
\ee 
Note also that \be \label{kmusy} \mu^{rs} (-\ka) = \mu^{sr}
(\ka)\, ,
 \ee where we have defined $\mu^{r+1, s+1} = \mu^{rs}$
for superscripts mod $3$. The twist matrix $C$ has a well-defined
action on the eigenvectors \be \label{conv} \sum_{n=1}^{\infty}
C_{mn} v_n (\ka) = - v_m (- \ka) \,,  \qquad C_{mn} = (-1)^m
\delta_{mn}. \ee It follows from this equation that the even and
odd components of the eigenvectors satisfy the relations \be
\label{eodd} v_{2n+1} (-\ka) = v_{2n+1} (\ka) \,, \qquad v_{2n}
(-\ka) = - v_{2n} (\ka)\,.
 \ee

\medskip
\noindent
The eigenfunctions can be shown to be orthogonal and complete,
\ben
\label{compl}
&&\sum_{n=1}^{\infty} v_n (\ka_1) v_n (\ka_2) = \delta (\ka_1 - \ka_2)\,,
\nonumber  \\
&&\int_{-\infty}^{\infty} d \ka \, v_m (\ka) v_n (\ka) = \delta_{m,n}\, .
\een
Due to the relations~\refb{eodd}, we can  separate the even and
odd modes and write the completeness and orthogonality
relations~\refb{compl}
\ben
\label{compl2}
&& 2 \sum_{n=1}^{\infty} v_{2n-1} (\ka_1) v_{2n-1} (\ka_2)
= \delta (\ka_1 - \ka_2), \quad
2 \sum_{n=1}^{\infty} v_{2n} (\ka_1) v_{2n} (\ka_2)
= \delta (\ka_1 - \ka_2),
\nonumber  \\
&& 2 \int_{0}^{\infty} d \ka \, v_{2m} (\ka) v_{2n} (\ka) = \delta_{m,n},
\quad
2 \int_{0}^{\infty} d \ka \, v_{2m-1} (\ka) v_{2n-1} (\ka) = \delta_{m,n} \ .
\een
We stress that these equations hold only for $\ka_1,\ka_2>0$.
In terms of odd and even modes equation~\refb{eveqn} becomes
\ben
\label{eigen1}
& & \sum_{m=1}^{\infty} M^{rs}_{2n, 2m} v_{2m} (\ka) = \ha (\mu^{rs} (\ka)
+ \mu^{sr} (\ka) ) \,  v_{2n} (\ka), \\
& &  \sum_{m=1}^{\infty} M^{rs}_{2n, 2m-1} v_{2m-1} (\ka) =
\ha (\mu^{rs} (\ka)
- \mu^{sr} (\ka) ) \, v_{2n} (\ka), \\
& & \sum_{m=1}^{\infty} M^{rs}_{2n-1, 2m-1} v_{2m-1} (\ka) =
\ha (\mu^{rs} (\ka)
+ \mu^{sr} (\ka) ) \,  v_{2n-1} (\ka), \\
\label{eigen4}
& &  \sum_{m=1}^{\infty} M^{rs}_{2n-1, 2m} v_{2m} (\ka) =
\ha (\mu^{rs} (\ka)
- \mu^{sr} (\ka) ) \,  v_{2n-1} (\ka) \ .
\een

These properties allow us to introduce new oscillators whose mode
number is a continuous parameter. From~(\ref{compl2}), it is convenient
to introduce new
continuous oscillators $e_\ka$ and $o_\ka$ associated to the even
and odd mode sums respectively:
\ben \label{oeosc}
&& o_\ka^\dagger
 = - \sqrt{2}\,i \sum_{n=1}^\infty v_{2n-1} (\ka)
 a_{2n-1}^\dagger \, , \qquad
 e_\ka^\dagger =\,\,\,\sqrt{2} \,\,\sum_{n=1}^\infty \,\,v_{2n}
(\ka) \,a_{2n}^\dagger \, ,\\
\label{oeosc1}
& & a_{2n}^{\dagger} = \sqrt{2} \int_{0}^{\infty} d \ka
\, v_{2n} (\ka) \, e_{\ka}^{\dagger} \,, \qquad
 a_{2n-1}^{\dagger} =  \sqrt{2} i \int_{0}^{\infty} d \ka
\, v_{2n-1} (\ka) \, o_{\ka}^{\dagger} \, .
\een
We have introduced a factor  of $i$ in the first equation so that
they have the same  BPZ conjugation property
 \be bpz(o_\kappa) = -
o_\ka^\dagger\,, \qquad bpz(e_\kappa) = - e_\ka^\dagger \ .
 \ee
The new oscillators satisfy the commutation relations
 \be [\,
o_\ka , o_{\ka'}^\dagger ] = [e_\ka , e_{\ka'}^\dagger] = \delta
(\ka - \ka') \,, \qquad  [\, o_\ka , e_{\ka'}^\dagger ] = [e_\ka ,
o_{\ka'}^\dagger] = 0\,.
 \ee
Note  that for $\kappa=0$, the $e_{\kappa=0}$ oscillator vanishes,
and we only have $o_{\ka =0}$. This is because the $\kappa=0$
eigenvector is $C$-odd.
Note also that the change of basis from the discrete oscillators
into the continuous one \refb{oeosc} is a unitary transformation,
as can be checked using \refb{compl2}.

Using equations~\refb{eigen1}--\refb{oeosc1} and the completeness
relations~\refb{compl2}, after some algebra the three-string vertex
 \be |V_3\rangle = \exp \Bigl[ - {1\over
2}\sum_{r,s} \sum_{m,n} a_m^{r\dagger} (CM^{rs})_{mn}
a_n^{s\dagger}\Bigr]|0\rangle\,,
 \ee
can then be written in terms of $o_{\ka}, e_{\ka}$ basis as
 \ben \label{frmfin0}
 |V_3\rangle &=&\exp
\,\Biggl[\, \sum_{r,s} \int_{0}^\infty\hskip-5pt d\ka\,
\left\{ - {1 \ov 4} (\mu^{rs} (\ka) + \mu^{sr} (\ka) )
\left({e^{(r)}_{\ka}}^{\dagger} {e^{(s)}_{\ka}}^{\dagger}
+ {o^{(r)}_{\ka}}^{\dagger} {o^{(s)}_{\ka}}^{\dagger} \right) \right. \biggr.
\nonumber
\\
& & \biggl. \left. \qquad\qquad\quad
+ {i \ov 4} (\mu^{rs} - \mu^{sr}) \left({o^{(r)}_{\ka}}^{\dagger}
{e^{(s)}_{\ka}}^{\dagger}
- {e^{(r)}_{\ka}}^{\dagger} {o^{(s)}_{\ka}}^{\dagger} \right)
\right \} \biggr] |0\rangle\,.
\een
where $\vert 0 \rangle$ is now interpreted as a ``continuous tensor
product'' of oscillator ground states for $e_{\ka}, o_{\ka}$.
More explicitly, \refb{frmfin0} can be written as
 \ben \label{frmfin}
  |V_3\rangle &=&\exp
\,\Bigl[\, \int_{0}^\infty\hskip-5pt d\ka\,\Bigl\{  -{1\over 2}
\mu(\ka) \, \Bigl({o_\ka^{(1)}}^\dagger {o_\ka^{(1)}}^\dagger
+{e_\ka^{(1)}}^\dagger {e_\ka^{(1)}}^\dagger + \hbox{cyc} \Bigr)
\nonumber\\
&&\qquad\qquad\quad - {1\over 2} (\mu^{12}(\ka) + \mu^{21}(\ka))
\Bigl({o_\ka^{(1)}}^\dagger  {o_\ka^{(2)}}^\dagger
+ {e_\ka^{(1)}}^\dagger  {e_\ka^{(2)}}^\dagger+ \hbox{cyc} \Bigr) \\
&&\qquad\qquad\quad  
- {i\over 2} (\mu^{12}(\ka) - \mu^{21}(\ka))
\Bigl(   {e_\ka^{(1)}}^\dagger  {o_\ka^{(2)}}^\dagger
- {o_\ka^{(1)}}^\dagger  {e_\ka^{(2)}}^\dagger
+ \hbox{cyc} \Bigr)
\Bigr\}\Bigr] |0\rangle\,.\nonumber
\een
This is the desired form of the open string three-vertex.
In the next section we identify the Moyal structures explicitly.
The above matter vertex is restricted to the $p=0$ sector, and is appropriately
normalized.

\sectiono{Identification of Moyal structures} \label{s3}

In this section we use the results from the two  previous sections
to show that the star product indeed corresponds to Moyal products
for a continuous set of variables parametrized by $\ka \in [0,
\infty)$. We identify these variables and also give an
interpretation for the commuting mode encountered at $\ka=0$.

We have now completed our preliminary work finding both
the oscillator representation of the Moyal product in
section 2, eqn.\refb{canmoy}, and a presentation \refb{frmfin} of the
three-string vertex as a continuous tensor product of three-vertices.
Comparing the exponents in these vertices we see that an identification
\be
(a^\dagger, b^\dagger) \leftrightarrow (e_\ka^\dagger, o_\ka^\dagger)\,,
\ee
requires the conditions
\ben
\mu(\kappa) =  {-4 + \theta^2\over 12 + \theta^2} \,, \qquad
\mu^{12}(\ka) + \mu^{21}(\ka)=  {16\over 12+ \theta^2} \,, \quad
\mu^{12}(\ka) - \mu^{21}(\ka)  =
{8\theta\over 12+
\theta^2} \,.
\een
Note that the consistency condition $\mu + \mu^{12} + \mu^{21} =1$
is satisfied. Making use
of the explicit expressions
\refb{evexp} we find that the above equations are all satisfied
by choosing
\be
\label{thetamoyal}
\theta(\ka) = 2 \,\tanh \Bigl( {\pi\ka\over 4} \Bigr) \,.
\ee
This is the noncommutativity parameter associated to the Moyal
algebras. Note in particular that for $\ka=0$ where the pair of
oscillators collapses to just $o_{\kappa=0}$ we have a commutative
product.  The noncommutativity parameter grows as $\ka$ grows but
is bounded: $\theta(\kappa) \leq 2$.

Note that the Witten vertex differs from the
Moyal vertex by a c-number prefactor, as can be seen by
comparing~\refb{canmoy} with~\refb{frmfin}.
This means that, strictly speaking, the 
open string field theory product $*_W$ and the Moyal product $*$
are related by an overall (infinite) constant factor,
\be \label{twoprods}
f *_W g = C' f * g \, ,
\ee
which we will write out later in equation (\ref{cprime}).
Such a factor can of course be eliminated by a redefinition of the
string field $f \rightarrow C' f$.

\medskip
Having identified the noncommutative parameter we must now
determine what are the explicit forms of the coordinates that
enter the Moyal product. It is natural to expect that those will
be continuous coordinate modes $x(\ka)$ associated to the usual
position modes $x_n$ by relations similar to those in
\refb{oeosc}. As hinted at before, we will have to use even and
odd modes, and a Fourier transformation of the odd modes will be
necessary.

For this purpose we
consider the position and momentum eigenstates
\ben
\langle \vec X(\sigma) | &=&  \langle 0| \exp \Bigl( -
\vec x \cdot E^{-2}
\cdot
\vec x +\,\, 2i\, \vec a \cdot E^{-1} \cdot \vec x + {1\over 2} \,\vec
a\cdot
\vec a \Bigr) \,, \nonumber \\
\label{cohexp}
 \langle \vec P(\sigma) |&=&  \langle 0| \exp \Bigl( -
{1\over 4} \,\vec p \cdot E^{2} \cdot \vec p \,+ \, \,\vec a
\,\cdot E \,\cdot \vec p\,\,\,\, \,\,\, -\,{1\over 2}\,\, \,\vec
a\cdot \vec a \Bigr) \,,
\een
where
\be \hat x = {i\over 2} \, E
\cdot (a - a^\dagger) \,,\qquad \hat p =  E^{-1} \cdot (a +
a^\dagger)\,, \qquad E_{mn} = \sqrt{{2\over n}}\, \delta_{mn}\,.
\ee
 (This is working in the $\alpha'= 1/2$ convention). More
explicitly \be \label{xoscposc} \hat x_n = {i\over \sqrt{2n}} \,
\cdot (a_n - a_n^\dagger) \,,\qquad \hat p_n =  \sqrt{{n\over
2}}\, \,(a_n  + a_n^\dagger) \,,\qquad  n\geq 1\,, \ee and the
zero mode expressions are those in \refb{zmodepos}. The above go
along with the expansions \be \label{modeexp} \widehat X (\sigma)
= \hat x_0 + \sqrt{2} \sum_{n=1}^\infty  \hat x_n \, \cos
n\sigma\,,\qquad \pi\, \widehat P(\sigma) = \hat p_0 + \sqrt{2}
\sum_{n=1}^\infty \hat p_n \, \cos n\sigma\, .
 \ee
We must now
define new coordinate and momentum operators associated to the
oscillators $(e_\ka, e_\ka^\dagger)$ and $(o_\ka, o_\ka^\dagger)$
we have introduced. We let \be (\hat x_\ka, \hat q_\ka )
\leftrightarrow  (e_\ka, e_\ka^\dagger)\,, \qquad (\hat y_\ka,
\hat l_\ka ) \leftrightarrow (o_\ka, o_\ka^\dagger)\,, \ee using
the standard correspondences for zero-modes \refb{zmodepos}:
\ben
\label{newosctwo} \hat x_\ka &=& {i\over \sqrt{2}} \,(\, e_\ka
\,-\, e_\ka^\dagger)\,, \qquad \hat q_\ka = {1\over \sqrt{2}} \,(\,
e_\ka \,+\,  e_\ka^\dagger)\,, \nonumber
\\ \hat y_\ka &=& {i\over \sqrt{2}} ( o_\ka -  o_\ka^\dagger)\,,
\qquad \,\,\,\hat l_\ka = \,\,{1\over \sqrt{2}} \,\,( o_\ka +
o_\ka^\dagger)\,.
 \een
The coordinate operators $\hat x_\ka$ and
$\hat y_\ka$ are the ones for which the 3-string vertex has been
put in Moyal form. We must therefore now express them in terms of
the original operators $(\hat x_n, \hat p_n)$.  For this we use
\refb{oeosc} to first pass to $(a_n, a_n^\dagger)$ oscillators and
then \refb{xoscposc} to pass to $(\hat x_n, \hat p_n)$ operators.
One immediately finds
 \ben \label{thenewvar1}
 \hat x_\ka
&=&\,\,\sqrt{2} \,\,\sum_{n=1}^\infty  v_{2n}(\ka) \sqrt{2n}\, \,
\hat x_{2n}\,,\\
\label{thenewvar2}
 \hat y_\ka &=&-\sqrt{2} \sum_{n=1}^\infty
{v_{2n-1}(\ka)
\over\sqrt{2n-1}}  \,\, \hat p_{2n-1}\,, \\
\hat q_{\ka} & = & \sqrt{2} \sum_{n=1}^{\infty} {v_{2n}(\ka)
\over\sqrt{2n}}  \,\, \hat p_{2n}\,, \\
\hat l_\ka &=&\,\,\sqrt{2} \,\,\sum_{n=1}^\infty  v_{2n-1}(\ka)
\sqrt{2n-1}\, \, \hat x_{2n-1}\, .
 \een
Here we see that the Moyal coordinates, the eigenvalues of
$\hat x_\ka$ and $\hat y_\ka$,  are respectively  (i)  linear
combination of even conventional coordinates, and (ii) linear
combinations of odd momenta. Thus the Moyal structure is canonical
for the multiplication of string functionals where the odd
coordinates $x_{2n-1}$ are replaced by odd momenta $p_{2n-1}$ via
Fourier transformation. In other words
given string functionals
$\Psi_i(\{x_{2n}\}, \{ x_{2n-1} \})$ to be star multiplied,
use of Moyal product requires the transformations 
\be \label{coortr}
\Psi_i(\{x_{2n}\}, \{ x_{2n-1} \}) \quad \to \quad \widetilde
\Psi_i(\{x_{2n}\}, \{ p_{2n-1} \}) \quad \to \quad
\Psi_i^M( x(\kappa), y(\ka))\,,
 \ee
where $x(\ka)$ and $y(\ka)$ denote the eigenvalues of
$\hat x_\ka$ and $\hat y_\ka$ respectively.  The first arrow above
denotes Fourier transformation, and in the second arrow we just
reexpress the coordinate and momenta in terms of continuous
variables using \refb{thenewvar1} and \refb{thenewvar2}. In this final form, 
with
superscript $M$ for Moyal, the star product is just canonical
Moyal product with $\theta(\ka)$ for each $\ka \geq 0$, in a way that
will be made more precise in the next subsection.

\medskip
It is of interest to identify the nature of the commuting
coordinate associated to $\ka=0 \to \theta(\ka) =0$. Indeed for
$\ka=0$ the eigenvector $v(\ka=0)$ only has odd components, and
therefore, from \refb{thenewvar1} and ~\refb{thenewvar2} only $\hat
y_{\ka=0}$ survives. This is explicitly
\be \hat y_{\ka=0} = -
\sqrt{2} \sum_{n=1}^\infty {(-1)^{n+1}\over 2n-1} \hat p_{2n-1} =
-\sqrt{2} \Bigl( \hat p_1 - {1\over 3} \hat p_3 + {1\over 5} \hat
p_5 - \cdots \Bigr) \,.
\ee
We can identify this right hand side
using \refb{modeexp}. We note that the above linear combination
is, with $p_0 =0$, just the momentum carried by half the string
\be \hat y_{\ka=0} = - \sqrt{2} \int_0^{\pi/2} \pi \hat P(\sigma)
d\sigma = -\sqrt{2} \pi \hat P_L \,. \ee Our result therefore
states that $\hat P_L$ must behave as an ordinary commuting
coordinate as far as the string field vertex is concerned. Indeed,
this is the case for zero momentum functionals, as we explain now.
Since the open string vertex is defined by gluing right half
strings to left half strings it implements the following
conditions when multiplying string $(1)$ times string $(2)$  to
give string $(3)$: \be \hat P_R^{(3)} = - \hat P_L^{(1)}, \quad
\hat P_R^{(1)} = - \hat P_L^{(2)},\quad \hat P_R^{(2)} = - \hat
P_L^{(2)} \ee But for zero momentum string states $\hat P_R^{(1)}
= -\hat P_L^{(1)}$ and $\hat P_R^{(2)} = -\hat P_L^{(2)}$, and as
a consequence we have that the vertex requires \be \hat P_L^{(1)}
= \hat P_L^{(2)}=  \hat P_L^{(3)} \,, \ee which is the statement
that $\hat P_L$ eigenvalues  behave as a commuting {\it
coordinate} at the string vertex.

It is interesting to note that for zero momentum functionals
$\hat P_L$ is, up to a constant, the same as the operator $\hat P_L -
\hat
P_R$.   The connection of the $\ka=0$ eigenvector to $\hat P_L - \hat
P_R$ is
clear from the observation of  \cite{0111069} that
this eigenvector implies that the sliver functional is invariant under
opposite rigid displacements of the two halves of the string. More
precisely, the zero-momentum sliver is annihilated by
the action of  $\hat P_L - \hat P_R$. It follows that the commuting
coordinate $\hat y_{\ka=0} \sim \hat P_L$ vanishes on the sliver.
Of course, it will not vanish on general zero-momentum functionals.

\sectiono{Open string star as continuous Moyal products}
\label{s4}

We have shown that the  noncommutative algebra of open string
field theory can be identified with a continuous tensor product of
Heisenberg algebras labeled  by $\ka > 0$, with a noncommutative
parameter given in \refb{thetamoyal}:
\be \label{noncompa} 
 [x(\ka), y(\ka') ]_{*} = i \, 2 \tanh {\pi \ka \ov 4} 
 \, \delta (\ka- \ka') \, .
\ee
In this section  we shall give a  precise description of this
continuous tensor product in terms of a functional integral
defined  on the half line $\ka  > 0$.

We begin our analysis by looking at the inner product in
string field theory, and how it would look in the continuous
bases.
Formally under the procedure outlined in~\refb{coortr}, the inner
product in string field theory  becomes \ben \nonumber \int
\Psi_1 \ast \Psi_2 
& = &
\int D x(\sigma) \, \Psi_1 (x (\sigma)) \, \Psi_2 (x (\pi - \sigma)) \\
\nonumber & = & \int \left(\prod_{n=0}^{\infty} D x_n \right)\,
\Psi_1 (\{x_{2n}\}, \{x_{2n-1}\}) \, \Psi_2 (\{x_{2n}\}, \{- x_{2n-1}\}) \\
\nonumber & = & \int \left( \prod_{n=0}^{\infty} D x_{2n}  \right)
 \left( \prod_{m=1}^{\infty} D p_{2m-1}  \right) \,
\tilde \Psi_1 (\{x_{2n}\},\{p_{2m-1}\}) \,
\tilde \Psi_2 (\{x_{2n}\}, \{p_{2m-1}\}) \\
\label{formalin}
& = &  \int Dx(\kappa) Dy(\kappa) 
 \, \Psi_1^M \left[x(\ka) ,y(\ka)\right] \,
\Psi_2^M  \left[x(\ka) ,y(\ka)\right] \,, \een
where in the third
line we have performed a Fourier transform for the odd
coordinates, and in the last line we have passed to the continuous
variables representing the Moyal coordinates. Thus we see that 
the integration in string field
theory indeed reduces to the standard integration in the Moyal
coordinates. Note that the ``$-$'' sign before $x_{2n-1}$ in
$\Psi_2$ in the second line is important for this identification.
This result was foreordained as there is no other integral which
respects cyclicity of the trace.

The above formal expressions, however, are not really meaningful
until we have specified the precise integral measures $Dx_n,Dp_n$
and $Dx(\ka), Dy(\ka)$ used in each step. In particular, it is
desirable to have the measures so that the
transformations~\refb{thenewvar1} and~\refb{thenewvar2} are
orthogonal transformations.
Since the non-commutativity parameter
is a tensor under changes of coordinates, as long as
we restrict ourselves to orthogonal transformations
it is
meaningful to speak about the magnitude of the  Moyal parameter.

We now define the integral measure in each line of~\refb{formalin}
by requiring  the perturbative string ground state
 \be \label{wavegr}
  \Psi_{\rm
ground} = \exp\biggl[ - \half \sum_{n=1}^\infty n  x_{n}^2\, \biggr]\,,
\ee
 has unit norm in each step. Defined this way the
transformations~\refb{thenewvar1} and~\refb{thenewvar2} are then
guaranteed to be orthogonal. More explicitly, the measure $Dx_n$
in~\refb{formalin} is given by
 \be \label{mforx}
D x_n = \left({n\over \pi}\right)^{\ha} \, d x_n \ .
 \ee
This is the formal measure
for the metric
\be \label{metone}
 \Vert \delta X \Vert^2 \sim \int_0^\pi \delta X(\sigma)
\sqrt{-{d^2\over d\sigma^2}} \, \delta X(\sigma) \,.
\ee
By choosing our Fourier transform conventions,\footnote{We use
$\langle p|x\rangle = {1 \ov \sqrt{2}} e^{ipx}$.} the measure
$Dp_n$ is given by
 \be \label{mforp}
D p_n = \left({1 \ov \pi n}\right)^{\ha} \, dp_n \ .
 \ee
Equivalently we can determine~\refb{mforx} and~\refb{mforp} by
using the inner products of~\refb{cohexp}, e.g. one finds that
$$
\langle x'_n | x_n \rangle = \left( {\pi \ov n} \right)^{\ha} \,
\delta (x_n - x'_n) \ .
$$
Similarly, we define the measure $Dx(\ka), Dy(\ka)$ so that
\ben
\nonumber  \int Dx(\kappa)
Dy(\kappa)\exp\biggl[-\int_{0}^{\infty} d \ka \left(x(\ka)^2 +
y(\ka)^2 \right) + 2 i\int_{0}^{\infty} d \ka \left[ j_1(\ka)
x(\ka) + j_2(\ka) y(\ka)\right]
 \biggr] \\
 \label{mforka}
\qquad\qquad =  \exp\biggl[ -  \int_{0}^{\infty} d \ka \, \left( j_1(\ka)^2 +
j_2(\ka)^2 \right)  \biggr]\,. \qquad\qquad \qquad \qquad
\qquad\qquad \qquad \qquad
\een
Formally this is the
measure associated to the standard $L^2$ metric
\be \label{mettwo}
\Vert (\delta x, \delta y ) \Vert^2 \sim \int_0^\infty [(\delta x(\ka))^2 + (\delta y(\ka))^2] \ .
\ee
An additional point will be necessary. Consider the integral
\be
 \int Dx(\kappa) Dy(\kappa)
\exp\biggl[-\int_{0}^{\infty} d \ka \, G(\ka)\left(x(\ka)^2 +
y(\ka)^2 \right) \biggr]
= (\det G)^{-1/2}\,.
\ee
A natural way to define the above determinant is to write
\begin{equation} \label{defn}
\ln \det G =  \Tr \log \hat G =
\int_{0}^{\infty} d \mu(\ka) \,  \langle \ka |\log \hat G | \ka \rangle
= \int_{0}^{\infty} d \mu(\ka) \, \log G (\ka)   \,,
\end{equation}
where the eigenvalues  of the operator $\hat G$ are given by
$G(\ka)$, and $d\mu(\ka)$ is the spectral measure.
As discussed in \cite{0111281,0201015},
$$
d\mu(\ka) = \divergence\ d\ka ,
$$
where $L$ is the ``level regulator'' (i.e. one regulates
by approximating $K_1$ with an $L \times L$ matrix).
Thus
\be
\label{mnb}
 \int Dx(\kappa)
Dy(\kappa)\exp\biggl[-\int_{0}^{\infty} d \ka \, G(\ka)\left(x(\ka)^2 +
y(\ka)^2 \right) \biggr]
= \exp \Bigl( - {1\over 2}  \divergence
    \int_{0}^{\infty} d \ka \, \log G (\ka) \Bigr)\,.
\ee

\bigskip
In order to define the  functional Moyal product necessary to describe
open string star products, we now look into
the definition of wave functionals of the Moyal coordinates.
The ground state wave function~\refb{wavegr} can be written in
terms of
$(x(\ka), y(\ka))$ using the inverse relations of \refb{thenewvar1}
and~\refb{thenewvar2}.  The result is
 \ben
\label{nvf}
\Psi^M_{\rm ground} (x(\ka),y(\ka)) = \exp \biggl[ - \ha
\int_{0}^{\infty} d \ka \, \left(x(\ka)^2 + y(\ka)^2 \right)
\biggr] \ . \een
Note that this wave functional indeed has unit normalization, as
follows directly from \refb{mforka}.  We can do this more
systematically for general wavefunctionals by introducing
\be
\vec X (\ka)
= (x(\ka), y(\ka)), \quad \vec A_\ka = (e_{\ka}, o_{\ka}), \quad
 \Theta (\ka) =  \pmatrix{0 &  \theta (\ka)
\cr -\theta (\ka) & 0} \,, \ee with $\theta(\ka)$ given
by~\refb{thetamoyal}. 
The position eigenstate $\langle x(\ka), y(\ka)|$ is just the
state $\langle X|$ in \refb{cohexp} restricted to the even modes, times
the state $\langle P|$ also in \refb{cohexp}, restricted to the odd modes,
expressed
in terms of $(x(\ka), y(\ka))$ and $(e_\ka, o_\ka)$ using equations
 \refb{thenewvar1} and~\refb{thenewvar2}
and \refb{oeosc}.  The result is
 \be
\label{bnm}
 \langle x(\ka), y(\ka)|
= \langle 0|
 \exp \Biggl( - \int_{0}^{\infty} d
\ka \, \left[ {1\over 2} \vec X (\ka) \cdot \vec X (\ka) -
\sqrt{2} \, i \vec A_\ka \cdot \vec X (\ka) - {1\over 2} \vec
A_\ka  \cdot \vec A_\ka \right] \Biggr)\,.
 \ee
Wavefunctions associated to states are then easily obtained performing
the contraction
\be \Psi^M ( x(\ka), y(\ka)) = \langle x(\ka), y(\ka)|
\Psi \rangle \ . \ee
Note that indeed for the vacuum we recover \refb{nvf}.
It is easy to see that with the measure introduced
in~\refb{mforka} we have
\be
\label{csc} \int Dx(\ka) D y (\ka) \, \ \vert
x(\ka), y(\ka)\rangle \langle x(\ka), y(\ka)| = 1 \ .
\ee

\bigskip
As a warmup exercise, and a test of the definitions,
we calculate  the  wave function in Moyal coordinates
for the star product of the ground state with itself.
Making use of \refb{frmfin}, \refb{bnm} and \refb{mnb} we find:
\ben \label{twogsproduct}
\Psi^M_{|0\rangle *_W |0\rangle } (x(\ka), y(\ka))
&=& \langle x(\ka),
y(\ka)| 0\rangle *_W |0\rangle \nonumber\\
&=& \langle x(\ka),
y(\ka)|
\exp\Bigl[ -{1\over 2} \int_0^\infty \mu(\ka) \Bigl(o_\ka^\dagger o_\ka^\dagger
+e_\ka^\dagger e_\ka^\dagger\Bigr)  \Bigr] \nonumber\\
&=& \exp \biggl[ - \ha
 \int_{0}^{\infty} d \ka \, (1 + {\rm sech}{\pi \ka \ov 2} ) \left(
 x(\ka)^2 +  y(\ka)^2 \right) \biggr] \nonumber\\
&& \quad  \cdot \exp \biggl[ \divergence
 \int_{0}^{\infty} d \ka \, \log \left(1+ \ha {\rm sech}{\pi \ka \ov 2}
 \right) \biggr] \ .
 \een
Although the singular factor in this wave functional may seem strange, 
it diverges as $L\to \infty$ and the integral is finite,
it would also be present in the standard coordinates,
as it would arise from the
determinant $\det(1-CM)$. The norm of the above state is
\begin{equation} \label{norm1}
\exp\left[- \divergence  \int_{0}^{\infty} d \ka \,
\log
 \left({1+  {\rm sech}{\pi \ka \over 2} \over
(1+ {1 \over 2} {\rm sech}{\pi \ka \over 2})^2
} \right) \right]\,.
\end{equation}
On the other hand,
the norm of $|0\rangle *_W |0\rangle$ computed in the oscillator basis
is given by
$(\det (1 - M^2))^{-1/2}$
which is precisely eq~(\ref{norm1}) if we use~(\ref{defn}) and~\refb{evexp}.

\medskip
With the properly defined measures we can now write the continuous
Moyal star product in terms of the functional form.
Using
\be
\langle x(\ka), y(\ka)|  
\Psi_1 *_W \Psi_2 \rangle =
 \langle x(\ka), y(\ka)| \langle\Psi_1| \langle\Psi_2| V_3\rangle\,,
\ee
and introducing two complete sets of coordinates via \refb{csc}, we can write
 \ben
\label{moyalfunctional} 
(\Psi_1^M \ast_W \Psi_2^M)
\left(x(\ka),y(\ka)\right)  =  \int && \hskip-15pt Dx_1(\ka)
Dy_1 (\ka) Dx_2 (\ka) D y_2 (\ka) \,\nonumber \\
&& K \left( \vec X_1
(\ka),
\vec X_2 (\ka), \vec X (\ka) \right)\nonumber \\
 & &  \, \, \Psi_1^M \left(x_1 (\ka),  
y_1(\ka)\right) \, \Psi_2^M \left( x_2 (\ka), y_2 (\ka) \right)\,,
 \een
where  the kernel $K$ is given  by
\be
K \left( \vec X_1 , \vec X_2, \vec X_3  \right)
= \Bigl( \langle \vec X_1|\otimes \langle \vec X_2|\otimes\langle
\vec X_3|\Bigr) \, |V_3\rangle \,.
\ee
With $\langle \vec X|$ given
by~\refb{bnm} and $|V_3\rangle$ by~\refb{frmfin}, a calculation gives:
\be
\hskip-2pt K \left( \vec X_1 , \vec X_2, \vec X_3  \right)
=  \exp \biggl[ - 2i \int_{0}^{\infty} \hskip-5pt d \ka \,
\left (\vec X_1(\ka) - \vec X_3 (\ka) \right) \cdot \Theta^{-1}
(\ka) \cdot
\left(\vec X_2(\ka) - \vec X_3 (\ka) \right) + \divergence\,C \biggr]
\label{kerK}
\ee
where
$C$ is a constant given by
\be
 \label{normconstant} C =
D \int_{0}^{\infty} d \ka \, \log \left[{1 \ov 8 } 
\left ( { 12 + \theta^2(\ka)  
\ov \theta^2 (\ka)} \right) \right] = D \int_{0}^{\infty} d \ka \,
\log \left[{1 \ov 8 } \left (1 + 3 (\coth{\pi \ka \ov 4})^2
\right) \right] \ ,
\ee
where $D$ is the number of space-time dimensions.
The two equations above, together with \refb{moyalfunctional},
 give the explicit reformulation of the open  
string star product as a functional Moyal product. Our work
has been precise enough to keep track of infinite factors
that appear in this transcription and that are well-known
to arise in the usual discrete formulations.

Note that constant $C$ obtained above is different from that constant
which would arise by exponentiating the prefactor in the standard
definition~\refb{defkf} of the Moyal kernel -- this is due to
the difference in normalization between~\refb{canmoy}
and~\refb{frmfin}.  The  ratio of these two constants is
the constant $C'$ in (\ref{twoprods}) defining 
the overall multiplicative factor relating the Witten and Moyal products:
\be \label{cprime}
C' = \exp  \Bigl(\,\, \divergence D \int_0^\infty d\ka\ \log \Bigl[
\, {1\over 8} \,(12 + \theta^2(\ka)) \Bigr] \,\Bigr)  .
\ee
In the full theory, the factors $C$ and $C'$
will also obtain ghost contributions,
possibly changing $D$ to $D-26$.

\sectiono{Relation with half-string basis and Bars' work}
\label{s5}

In this section we discuss the relation of the $\ka$-basis to the
half string basis and compare our result with that of
Bars~\cite{0106157}.

The open string star product can be written in a functional form in terms
of overlap of half strings \be
 \left(\Psi_1 *  \Psi_2\right) [x_L (\sigma), x_R (\sigma)]
        \equiv \int \prod_{{0} \leq \sigma \leq {\pi\over 2}}
Dz(\sigma) \,  \Psi_1 [x_L(\sigma), z (\sigma)] \,
   \Psi_2 [z (\sigma), x_R(\sigma)]\,,
\label{mult}
 \ee 
where the open string halves are expanded as:
\begin{eqnarray} \label{eq:multi}
&& x_L(\sigma)  =  x(\sigma)  = l_{0}+\sqrt{2}\sum_{m=1}^{\infty}\
l_{2m}\cos(2m\sigma), \quad \qquad   {0} \leq \sigma \leq {\pi\over 2}\,, \\
& & x_R(\sigma)  = x(\pi - \sigma)  =
r_{0}+\sqrt{2}\sum_{m=1}^{\infty}\ r_{2m}\cos(2m\sigma), \quad {0}
\leq \sigma \leq {\pi\over 2} \ .
\end{eqnarray}
and
\begin{equation} \label{hastrb}
l_{2m}=x_{2m}+\sum_{n=1}^{\infty}\mathcal{T}_{2m,2n-1}x_{2n-1},\qquad
r_{2m}=x_{2m}-\sum_{n=1}^{\infty}\mathcal{T}_{2m,2n-1}x_{2n-1}. \label{even}
\end{equation}
with
 \be
  \CT_{2n,2m-1} ={4 \ov \pi} \int_0^{\pi \ov 2} d \sigma \,
 \cos 2 n \sigma \, \cos(2m-1) \sigma \ .
  \ee

We have seen in the last section that the open string star product in
the operator form  can be written (up to an infinite constant) 
as the canonical Moyal product
in terms of coordinates $(x_{\ka}, y_{\ka})$ with a continuous
label $\ka > 0$. Here we make a direct connection
with~\refb{mult}.

By inverting~\refb{thenewvar1},~\refb{thenewvar2} and
using~\refb{noncompa} we can go back to the standard Fock  space
basis and interpret the open string star product as the Moyal product
in terms of coordinates $(\{x_{2n}\}, \{p_{2m-1}\})$ \be
\label{xpcom} [x_{2n},p_{2m-1}]_{\ast} = i \Theta^{2n,2m-1}\,, \qquad
n,m\geq 1\,,
 \ee
where from
equations~\refb{noncompa},~\refb{thenewvar1},~\refb{thenewvar2},
the noncommutative parameter $\Theta$ is given by
\be \label{thetafock}  \Theta^{2n,2m-1}= -2 \sqrt{2m-1\over 2n}
\int_{-\infty}^{+\infty} v_{2m-1}(\ka) v_{2n}(\ka) \tanh(\pi\ka/4)
d\ka\,. \ee
Note that here $p_{2m-1}$ play the role of ``coordinates'' while
$x_{2m-1}$ are interpreted as the corresponding ``momenta''.
Using ~\refb{xpcom} and recalling that $x_{2m-1}$ are the momenta
conjugate to the noncommutative coordinates $p_{2m-1}$, we see
that~\refb{mult}--\refb{even} precisely have the form of the Moyal
product in the mixed coordinate and momentum
basis~\refb{theblo}--\refb{moymix} if we identify~\footnote{In
this paper we ignore the zero mode part, i.e $n=0$ elements of the
matrix.}
\be 
\label{barstwo}  
\ha \Theta^{2n,2m-1} \quad
=  \quad  {\cal T}_{2n,2m-1}\qquad n,m\geq 1 \ .
\ee

We will now prove~\refb{barstwo} by considering generating
functions. We define
\ben \nonumber  \tilde \CB(z,w)& : = &
\ha  \sum_{n,m\geq 1} z^{2n} w^{2m-1}   \Theta^{2n,2m-1}\\ 
\label{genthettilde} & = & - \half  {w\over 1+w^2} \,
\int_{-\infty}^{+\infty} d\ka \,  {\tanh(\pi\ka/4)\over \sinh(\pi
\ka/2)} \cosh \ka v (1 - \cosh \ka u)\,,
\een
with $u=\tan^{-1} z$ and $v = \tan^{-1} w$.
In the second line above we have inserted the
generating function~\refb{genaf} for $v_n$ and the explicit
expression for $\theta(\ka)$. Similarly we consider
\ben \nonumber
 \CB(z,w) & := & \sum_{n,m\geq 1}^\infty z^{2n}
w^{2m-1} \CT_{2n,2m-1} \\
\label{barsfunc} & = &  {z^2 w \over 4\pi i} \oint {d\xi\over
\xi^{1/2} } \Biggl({\xi \over 1-z^2\xi} + {1\over \xi - z^2}
\Biggr)\Biggl( {1\over 1-w^2\xi } + {1\over \xi - w^2} \Biggr)\,,
 \een
where the integral extends from $\xi=e^{-i\pi}$ along the unit
circle to $\xi = e^{+i\pi}$. It can be shown by  evaluating the
integrals~\refb{genthettilde} and~\refb{barsfunc}  that indeed \be
\tilde B(z,w) = B(z,w) = {i\over \pi} {(1-w^2)z\over
(z^2-w^2)(1-z^2 w^2)} \Biggl[ z(1+w^2)\log\bigl({1+iw  \over
1-iw}\bigr) - w(1+z^2) \log\bigl({1+iz  \over 1-iz}\bigr) \Biggr]
 \ee The first integral~\refb{genthettilde} is easily done by
closing the contour in the upper half plane. To do the second
integral we deform the contour, picking up the residues at $\xi =
z^2, w^2$ to an integral around the cut from $0$ to $-1$. To do
the integral around the cut one needs:
\be \label{barsfuncii} \int_0^1 {dx\over x^{1/2}} {1\over 1+a^2 x}
= {1\over i a} \log({1+i a\over 1-ia}) \ . \ee
This completes the verification of \refb{barstwo}.

Finally let us compare the noncommutativity
parameter~\refb{thetafock} with that given by Bars
in~\cite{0106157}, eq.~(33):
\be \label{barscomm}
[x^{\Bars~\mu}_{2m},p^{\Bars~\nu}_{2n}] = i\eta^{\mu\nu} \delta_{m,n}\,\, .
\ee
To compare with
our results we note that $x_{2n}^{\rm here} = x_{2n}^{\Bars}$
and from \cite{0106157}, eq.~(10):  
\be \label{barsone} p_{2m-1}^{\rm here } = 2 \sum_{n=1}^\infty
p_{2n}^{\Bars} {\cal T}_{2n,2m-1}\,, \ee and therefore
equation~\refb{barstwo} indeed shows they are equivalent.

\smallskip

While this is a satisfying result, it is also somewhat puzzling.
We have found a continous spectrum of noncommutativity parameters,
including $\theta(\kappa=0)=0$, in contrast to Bars' result (\ref{barscomm}).
Since, as we have just shown,  ${\cal T}$ is essentially
the same as $\Theta$, and since the latter matrix has a
zeromode, it follows that ${\cal T}$ has a zeromode,
and hence we cannot invert the relation (\ref{barsone}).

It might sound ridiculous to say that ${\cal T}$ is not
invertible, since one can introduce the explicit
matrix ${\cal R}$ given by
\be \label{arrmat}
{\cal R }_{2m-1,2n} = {\cal T}_{2n,2m-1} - (-1)^n {\cal T}_{0,2m-1}\,,
\ee
which satisfies \footnote{Essentially equivalent identities occur in the
paper of Gross and Taylor \cite{GTI} concerning the construction of
orthogonal transformations from their $X_{eo}$ and $X_{oe}$.}
\be \label{invt}
({\cal R}{\cal T})_{2m-1,2k-1} = \delta_{m,k}\,,
\qquad
({\cal T}{\cal R})_{2n,2k} = \delta_{n,k}\,.
\ee
The sums implicit in (\ref{invt}) are absolutely convergent;
doesn't this mean that ${\cal T}$ is invertible?
Not necessarily, because  the spectrum of an operator depends
on the linear space on which it is defined. For example, while
${\cal T}$ is invertible on the Hilbert space $\ell_2$
of square-integrable sequences it is not invertible on the
Banach space $\ell_{\infty}$ of bounded sequences. (The
eigenmodes $v_n(\ka)$ are  elements of this latter space.)
Which linear space one should use in formulating
string field theory
is a matter of physical definition. Since the zero mode
of ${\cal T}$ appears to be physically meaningful, and
is moreover important in the $\alpha' \to 0$ limit
\cite{0111069} we believe we should take seriously
the zero mode of ${\cal T}$, and not make redefinitions which
obscure it.  Further evidence for this
point of view will appear in the next section.\footnote{We should mention that
Bars and Matsuo \cite{0202030}
have also addressed subtleties associated with this zero mode,
and have suggested that these be thought of in terms of explicit
``associativity anomalies.''}

\sectiono{String field theory as NC field theory} \label{s6}
The results in previous sections allow us to rewrite
the cubic interaction as
\be
\int \Psi * \Psi * \Psi ,
\ee
where $*$ is now the continuous tensor product
of Moyal products, and $\int$ is the standard
integration in terms of the Moyal coordinates.
\footnote{We expect that inclusion of the ghosts into
this discussion should be possible.}
We would now like to write the entire   matter part of the
bosonic string field theory action,
\be
S = \int {1\over 2} \Psi  (L_0-1) \Psi + {1\over 3}\Psi * \Psi * \Psi ,
\ee
as a noncommutative field theory action.
Hence we now turn to a detailed discussion of the
kinetic term.

We are going to find difficulties in writing
a well-defined kinetic term when we use variables
in which the $*$-product is simple, therefore
we begin with a general discussion. The string
field wave function $\Psi$ can be written as a
function of infinitely many variables in many
ways. The standard way is to use the variables
$x_n$ in the Fourier expansion of $X(\sigma)$.
However, we have seen that one may wish to
Fourier transform some variables and take
nontrivial linear combinations of the resulting
coordinates.  Therefore,
let us simply assume that the
string field is a function of some infinite
collection of variables $z^i$ such that the kinetic
operator can be written as
\be \label{someoper}
L_0 = \sum_{ij} g^{ij} a_i^\dagger a_j\,.
\ee
($i,j$ here can be continuous, as with the $\kappa$ basis, or
discrete). We assume the oscillators $a_i$ act as:
\ben
a_i\Psi& =& f_i * \Psi + \Psi * g_i \,,\nonumber \\
a_i^\dagger \Psi& =& f_i^* * \Psi + \Psi * g_i^*\,,
\een
for some elements $f_i, g_i$ of the algebra.
It is then straightforward to check that the kinetic term can be
written as:
\be
\int \Psi * L_0  \Psi =
 \int g^{ij} {\rm Re} [f_i,\Psi] [g^*_j,\Psi] + \int \Psi * \Psi * V,
\ee
with
\be
V = \ha g^{ij} \left( \{f^*_i+g^*_i,f_j+g_j\}_*
  + [f^*_i,f_j]_* + [g_i,g^*_j]_* \right)\,,
\ee
where all products are star products.

We are now going to examine three choices of basis in which the star
product is reasonably simple. We will find difficulties with all three
bases. Indeed, the generality of the difficulties can be seen by
looking at invariants under linear changes of coordinate, such as the
spectrum of linear operators.

For our first choice of coordinates, we take
\be \label{ourcoords}
z^\mu = \pmatrix{ x_{2n} & p_{2m-1}} .
\ee
Here $\mu$ is an index running over positive even, then positive odd integers.
In terms of these coordinates the standard string field kinetic term is
given by
\ben
\label{ellnought}
L_0 &=& \sum_{n\geq 1} n a_n^\dagger a_n \nonumber \\
&=& \half \sum (p_n^2 + n^2 x_n^2) \nonumber\\
& =& \half \sum \bigl( - {\p^2\over \p x_{2n}^2} + (2n)^2 x_{2n}^2\bigr)
 + \half \sum \bigl(
 - (2m-1)^2 {\p^2\over \p p_{2m-1}^2} +  p_{2m-1}^2\bigr)\nonumber\\
& =& \half \sum_\mu \biggl[
 - g_{(1)}^{\mu\nu}\p_\mu \p_\nu + (g_{(2)})_{\mu\nu} z^\mu z^\nu\biggr]\,,
\een
where we have introduced the metrics:
\ben
g_{(1)}^{\mu\nu} & =\delta_{2n,2n'} \oplus  (2m-1)^2 \delta_{2m-1,2m'-1}\,, \\
(g_{(2)})_{\mu\nu} & =  (2n)^2 \delta_{2n,2n'} \oplus \delta_{2m-1,2m'-1}\,.
\een

Quite generally, given an invertible noncommutativity parameter
\be
[z^\mu, z^\nu] = i \Theta^{\mu\nu} ,
\ee
one can rewrite a derivative as a star commutator,
\be \label{subderiv}
\p_\mu \Psi = -i \Theta^{-1}_{\mu\nu}[ z^\nu, \Psi]_*
\ee
Since the string field theory trace $\int $ is the same as the
noncommutative field theory trace $\int$, we can use  cyclicity of the
trace, (and the fact that the metrics and $\Theta$ are constant, as
functions of $z^\mu$) to obtain
\ben \label{noncomactone}
\int \Psi L_0 \Psi  =
 \half \int   [z^{\mu}, \Psi]_* Q_{\mu\nu}  [z^{\nu},\Psi]_*
+( g_{(2)})_{\mu\nu}( z^\mu z^\nu) *( \Psi^2)\,,
\een
where
\ben
Q_{\mu\nu}
 = \Theta^{-1}_{\mu\rho} g_{(1)}^{\rho\lambda} \Theta^{-1}_{\lambda\nu}
 =\sum_n \tilde \Theta^{-1}_{\mu,2n} \tilde \Theta^{-1}_{2n,\nu} +
\sum_m \tilde \Theta^{-1}_{\mu,2m-1} (2m-1)^2 \tilde \Theta^{-1}_{2m-1,\nu}\,,
\een
and all products are commutative products unless explicitly indicated as
*-products.

This presentation of the action is very close to that for a generic
noncommutative field theory \cite{0106048}, with the slight difference
that the potential explicitly breaks translation symmetry.  One might
therefore hope to adapt results such as the existence of solitonic solutions
very directly to string field theory, thus justifying the
identification of D-branes with string field theory solitons.

The form (\ref{noncomactone}) is actually somewhat deceptive for two
reasons.  First, as we have noted, $\Theta$ has a zero mode, so it
does not have an inverse.  Of course this is a standard situation in
noncommutative field theory, and one can easily enough deal with it by
not making the substitution (\ref{subderiv}) for commuting
coordinates.  What is potentially more problematic here is that
$\Theta$ has a continuous spectrum around zero.  This opens the
possibilities that one will not be able to separate out the zero mode
usefully, or that the rewriting (\ref{subderiv}) might lead to
divergences even for the noncommuting coordinates with arbitrarily
small $\theta$.

A more fundamental difficulty, however, is that the products of
the operators $\Theta$, $g_{(1)}$ and $g_{(2)}$ which arise in evaluating
the kinetic term derived from (\ref{noncomactone}), are ill defined.
Consider for example the potential term; evaluating it on a generic
string functional will involve acting with operators such as
\be \label{potoperator}
\Theta g_{(2)} \Theta g_{(2)} .
\ee
Unfortunately, this operator does not
really exist -- and this is a basis-independent statement.

The problem becomes clear when one attempts to
express $L_0$ in the $\kappa$-basis as an integral kernel.
That is, we search for a function $K(\ka,\ka')$ such that
\be \label{deflzeroker}
[L_0, a_\ka] = \int_{-\infty}^{+\infty} d\ka' K(\ka,\ka') a_{\ka'}\,.
\ee
Unfortunately the formal solution to (\ref{deflzeroker}),
\be \label{formser}
K(\ka,\ka')  = \sum n v_n(\ka) v_n(\ka') ,
\ee
does not exist because the series does not converge. This is
apparent from the large $n$, fixed $\ka$, asymptotics of
$v_n(\ka)$ ($N$ below is given by~\refb{defnorm}):
\def\cn{{\cal N}}
\ben \label{asymptos}
\sqrt{n+1}v_{n+1}(\ka)
 & \sim & {(-1)^{n/2} \over N(\ka)^{1/2}}  {\rm Re}\Biggl[
{ (2n)^{i\ka/2}\over \Gamma(1+i \half \ka)} \Biggr] \qquad\,\,
\,\,\, n\quad even\,, \nonumber \\
& \sim &{(-1)^{(n+1)/2} \over N(\ka)^{1/2}}  {\rm Im}\Biggl[
{ (2n)^{i\ka/2}\over \Gamma(1+i \half \ka)} \Biggr] \qquad n\quad odd\,,
\een
and hence the series (\ref{formser}) does not converge.
One can check that direct evaluation of (\ref{potoperator}) in the
first ($n$) basis leads to the expressions containing
the same nonconvergent series (\ref{formser}).

One might imagine that a kernel satisfying (\ref{deflzeroker}) still
exists, but that it cannot be obtained by doing the sum (\ref{formser}).
This hypothesis can be tested by considering the regularized operator
$L_0^q: = \sum q^n a_n^\dagger a_n $, where
$0< q< 1$. Then one can indeed find a well-defined
kernel satisfying:
\be \label{defkernl}
[L_0^q , a_\ka] = \int d\ka' K_q(\ka,\ka') a_{\ka'}\,,
\ee
where $L_0^q$ is a regularized version of $L_0$.
However, $K_q(\ka,\ka')$ does not have a well-defined
$q\to 1$ limit, as is already shown by
the identity
\be \label{mixer} K_q(\ka,0) = {1\over \sqrt{4\pi N}} {q\over
1-q^2} \biggl[ \Bigl({1-q\over 1+q}\Bigr)^{i \ka/2} +
\Bigl({1+q\over 1-q}\Bigr)^{i \ka/2} \biggr]\,. \ee
A similar, but more elaborate, formula exists
for the full kernel $K_q(\ka,\ka')$, and shows
that there is no well-defined $q\to 1$ limit
for all values of $\ka,\ka'$.

Having found that our bases are problematic,
let us turn to the basis advocated in
\cite{0106157}. Rather than using
$x_{2m},p_{2m-1}$ we now use $\bar y, y_{2m}$
and $p_{2m}^{\rm Bars}$, where
$\bar y = x_0 + \sqrt{2}\sum (-1)^n x_{2n}$
and $y_{2n}=x_{2n}$. Since
\be
{d\over dx_0}  = {d\over d\bar y}\,,
\ee
\be
{d\over dx_{2n}} =
{d\over dy_{2n}} + \sqrt{2}(-1)^n
{d\over d\bar y}\,,
\ee
the zeromomentum sector with respect to $x_0$ coincides with
that with respect to $\bar y$. We restrict ourselves to this
sector since nonzero momentum with respect to $\bar y$, with
finite momentum with respect to $y_{2m}$ appears to be problematic.

Now,  using the identities
\be \label{curlytee}
\sum \CT_{2n,2m-1} \CT_{2n',2m-1} = \delta_{n,n'}\,,
\ee
valid for $n,n' > 0$, and
\be \label{badident}
\sum_{m\geq 1} (2m-1)^2 \RR_{2m-1,2n} \RR_{2m-1,2n'}
= (2n)^2 \delta_{n,n'}\,,
\ee
we convert $L_0$  to
\ben
L_0
 = \half \sum \bigl( - {\p^2\over \p x_{2n}^2} + (2n)^2 x_{2n}^2\bigr)
 + \half \sum \bigl( - m^2 {\p^2\over \p( p_{2m}^{\rm Bars})^2}
 +  ( p_{2m}^{\rm Bars})^2\bigr)\,.
\een
Evidently, we have not reproduced the standard perturbative string spectrum!
Instead of a single tower of oscillators with one oscillator for each
integral frequency, we have one oscillator for every odd frequency and
two oscillators for every even frequency. What has happened?

Once again, one can trace the difficulty back to the fact that the
spectrum of an operator is sensitive to the linear space on which it
is defined.  As noted in \cite{0111069},
identities such as (\ref{badident}) imply, naively,
that one can change the set of eigenvalues of a diagonal matrix by an
invertible transformation. This absurdity results from
neglecting subtleties in dealing with unbounded operators
(e.g. see \cite{ReedSimon}).
Evidently, such niceties of functional analysis
are  relevant for string field theory.

\sectiono{Discussion and new directions} \label{s7}

We can summarize the main conclusions of our discussion as follows.

\begin{itemize}

\item
We have a Moyal description of the star product which
arises unambiguously from the standard mode basis
by a unitary transformation, equation (\ref{oeosc}),
implemented by a well-behaved infinite dimensional matrix.

\item
It leads to a tensor product of algebras with a continuous spectrum
   of $\theta$'s including zero. While a finite number of commutative
   modes can be incorporated into an NC field theory framework by treating
   them specially, a continuous spectrum around zero is something new
and poses interesting questions.

\item The
theta spectrum is bounded above.  We believe this
    is physically meaningful, since we did a well-defined unitary change
   of basis.  It means that the open string star noncommutativity has a
definite  ``maximal range" in the usual NC field theory sense, where space
nonlocality is related to the momentum as $\Delta x = \theta p.$
   This seems  most intriguing to us and
   its physical implications deserve more thought.

\end{itemize}

We should remind the reader that the inner product with respect to
which our transformation is unitary is equation (\ref{metone}); this
is the quadratic form which (for example) appears in the ground state
(or any Fock state) wave functional and thus can be said to govern
the $\alpha'$ fluctuations of a free string.  The boundedness of $\theta(\ka)$
is therefore the statement that the noncommutativity of the interaction
term is comparable to (or less than, for modes with small $\ka$) the
quantum uncertainty of the free string. 

\medskip
We also note that the precise relation between the Witten and Moyal
products (\ref{twoprods}) involved an overall multiplicative factor,
(\ref{cprime}).  Of course, we computed this only for the matter sector
and the ghosts will clearly modify this result.  
Since ultimately we are interested in wavefunctionals
for which the open string star product is finite and thus
define the open string algebra, this factor
must be taken seriously in passing to the Moyal product formulation.

\medskip
An important extension of the present work would be to
discuss the reformulation of the star product in terms
of the Moyal product for arbitrary string functionals, that
is, relaxing the zero momentum condition imposed in the 
present paper. The spectrum of the
requisite Neumann matrices is not completely known, but it is
expected to include a continuous set of eigenvalues from
$[-1/3, 0)$ with the $(-1/3)$ eigenvalue corresponding to the
$C$-even eigenvector discussed in \cite{0111069,unp}. It is clear
that associated to this eigenvector, the corresponding commuting
coordinate is the string midpoint.  Presumably there is no
$C$-odd eigenvector for this eigenvalue, so the results might
take a form rather similar to those in this paper. On the other
hand it is known from numerical experiments \cite{unp} that the
spectrum also includes a pair of degenerate eigenvectors
with eigenvalue in the interval $(0,1)$. The interpretation
of the associated Moyal coordinates could be of interest.

\medskip
One of the more interesting problems left open by the above discussion
is the apparent conflict between finding a simple description of the
kinetic term in the string field action and the cubic interaction
term. In the standard presentation, the kinetic term is simple, but
the star product is complicated.  In the present paper we have given
another presentation of the string field theory action which makes the
star product simple, but in which it is difficult to write a
well-defined kinetic term.  This conflict seems to persist among
various choices of coordinates for the string field, though we have by
no means ruled out the existence of coordinates which fix this
problem, perhaps obtained by more complicated operations such as
additional Fourier transforms. 
It seems important to understand this difficulty better.  Since there is
no difficulty in defining the regularized operator $L_0^q$,
computations can be done in this presentation, by taking the
$q\rightarrow 1$ limit only at the end of the computation.  Evidently the
ill-defined quantities discussed in section 7 cancel out of physical
results.

\medskip
There is some evidence both for and against the idea that the star product
on physical states really is singular, in the sense of not defining an
algebra of bounded operators.  In \cite{0111069}, it was argued that
the squeezed state defining the star product in the matter sector
involves a quadratic form
which is not Hilbert-Schmidt and therefore the open string star product
does not give states of finite norm.
Here we confirmed this in Moyal coordinates, in equation
(\ref{twogsproduct}).
On the other hand, it must be
admitted that many sensible computations have been done using level
truncation and the unregularized Witten product.  It may also
be premature to discuss these issues before taking proper account of
ghosts and the BRST operator.
If it were to turn out that the star product had to be regulated to
get a sensible formalism, one would lose the associativity of the algebra
(even though the $\ka=0$ mode was being treated properly -- so this
is different from the nonassociativity of \cite{0202030}).
In place of associativity, one would have 
$A_{\infty}$ structure \cite{Stash},  
which also has a well developed theory, though not nearly so well
developed as conventional operator algebras. Such algebras
have featured in constructions of open string field theory
\cite{9705038,9705241}. 

\medskip
The presentation of the star algebra in this paper
is an important step in formulating rigorously what should be
meant by the K-theory of the string field theory algebra.
It is widely expected that D-branes should be topologically
classified, in the context of string field theory,  in terms
of the $K$-theory of some
noncommutative algebra related to a vertex operator algebra.
Certainly the  K-homology
of $C^*$ algebras does appear to be physically
relevant to the topological
classification of D-branes in the long-distance
limit of the theory. Nevertheless, the precise
definition of the K-theory of the entire string field theory
algebra has not yet been given. Writing the algebra as a
continuous product of standard Heisenberg algebras should open
the way to giving such a precise definition.
Nevertheless, there is much
nontrivial work left to do. The resolution
of the difficulties we have pointed out with the kinetic
term will have an impact on how this definition is carried out.
Moreover, assigning rigorous meaning to the continuous
tensor product of $C^*$ algebras might well be nontrivial.

\bigskip

We would like to thank I. Bars for a conversation.
Comments by R. Dijkgraaf, L. Rastelli and A. Sen are also acknowledged.
This work was supported by DOE grant DE-FG02-96ER40959.
B.~Zwiebach would like to acknowledge the hospitality of
the Rutgers Physics Department.  The work of  B.Z. 
was supported in part
by DOE contract \#DE-FC02-94ER40818.

\vfill
\break

\end{document}